# Anisotropic Dielectric Function of Graphite Probed by Far- and Near-Field Spectroscopies

*Adilet Toksumakov[1#] [\*], Georgy Ermolaev[1#], Dmitriy Grudinin[1#], Aleksandr Slavich[1], Nikolay Pak[1], Gleb Tikhonowski[1], Andrey Vyshnevyy[1], Gleb Tselikov[1], Aleksey Arsenin[1], Valentyn Volkov[1\*]*

[1]Emerging Technologies Research Center, XPANCEO, Internet City, Emmay Tower, Dubai, United Arab Emirates

[#]These authors contributed equally to this work

\*Correspondence should be addressed to the e-mail: vsv@xpanceo.com, adilet.toksumakov@xpanceo.com

## Abstract

**Graphite is a cornerstone material for revolutionary technologies, from energy storage to the entire field of two-dimensional materials. Despite its foundational role, the predictive power required for engineering emergent optical behavior in van der Waals heterostructures is severely constrained by persistent discrepancies in reported optical constants. We resolve this long-standing ambiguity by deploying a multi-modal approach that synergizes far-field spectroscopic ellipsometry with nanoscale near-field optical probing (s-SNOM) and micro-reflectance spectroscopy. We have established a new, self-consistent set of optical constants (n and k) for both in-plane and out-of-plane crystallographic directions across the ultraviolet-to-near-infrared spectrum. This work presents a unified set of optical constants that addresses inconsistencies in the existing literature. By establishing this definitive reference, we provide the essential foundation for the quantitative modeling and engineering of light-matter interactions in the evolving landscape of carbon-based nanophotonics.**

## Introduction

Graphite, a crystalline allotrope of carbon, is defined by its highly anisotropic hexagonal layered structure composed of $sp^2$-hybridized carbon atoms arranged in two-dimensional graphene sheets[1,2]. These sheets are held together by weak van der Waals forces, which gives rise to a class of materials known as van der Waals (vdW) materials. The unique structure leads to a pronounced contrast in physical properties between the in-plane and out-of-plane directions. Such intrinsic anisotropy governs many of graphite's physical behaviors and is central to its technological relevance across diverse domains, ranging from classical applications such as lubricants and electrodes to modern uses in nanoelectronics, photonics, and optoelectronics — particularly following the advent of graphene[3] and other two-dimensional (2D) materials[4,5]. The advent of 2D materials has placed vdW materials at the forefront of modern materials science, opening new frontiers in nanophotonics and optoelectronics due to their unique properties, such as high refractive indices and strong optical anisotropy[6–10].

Among the key descriptors of a material's interaction with electromagnetic radiation are its complex optical constants — the refractive index (n) and extinction coefficient (k). These quantities provide direct insight into the material's electronic band structure and dictate optical phenomena such as absorption, reflection, and transmission. The fundamental linear optical properties of bulk graphite have been well-established since the seminal work of E. A. Taft and H. R. Philipp in 1965[11] which characterized its dielectric function by the distinct contributions of π and σ electron transitions and identified a strong ultraviolet π-plasmon resonance[12,13].

Since those early studies, the focus of the research community has substantially shifted. Attention has been directed primarily towards the distinct optical effects observed in graphene, graphite's monolayer. Graphene exhibits a unique, frequency-independent optical absorption of ~2.3%[14,15] and allows for dynamic tuning of its properties via electrostatic gating[16–19], making it a star material for applications in THz plasmonics and nonlinear optics[20,21]. Similarly, other vdW materials like transition metal dichalcogenides (TMDCs) have garnered significant attention because they undergo a transition from an indirect bandgap in bulk to a direct bandgap as monolayers, enabling highly efficient light emission and unique spin-valley physics[22–28].

This intense focus on single– and few–layer systems has created a distinct research gap: while the linear optics of bulk graphite are considered classic knowledge, its properties are often overlooked in the context of modern nanophotonics. There remains a critical need to bridge the gap between the foundational understanding of bulk graphite and the cutting-edge phenomena observed in its 2D analogues. Despite decades of research, a comprehensive and consistent dataset of graphite's anisotropic optical constants across a broad spectral range remains lacking, with considerable discrepancies in the literature[29–34]. These inconsistencies are often attributed to differences in experimental techniques, sample quality, and data modeling strategies that do not sufficiently account for graphite's inherent uniaxial anisotropy.

In this study, we revisit bulk graphite to report a comprehensive, high-accuracy determination of its anisotropic optical constants over a wide spectral range (250–1700 nm) using a multi-technique approach. By integrating spectroscopic ellipsometry with Raman spectroscopy for structural validation and near-field scanning optical microscopy (SNOM) and direct reflectance measurements for verification of optical constants, we provide a definitive reference dataset. Our work aims to resolve longstanding inconsistencies and establish a reliable foundation for modeling graphite-based components in advanced photonic, plasmonic, and optoelectronic systems, thereby reaffirming the relevance of this foundational material in the age of 2D photonics.

**Results**

We began our investigation by characterizing the structural and vibrational properties of the bulk graphite source material. As a van der Waals crystal, graphite possesses a hexagonal layered structure (space group $P6_3/mmc$) with pronounced anisotropy, as depicted in Fig.1a. The high crystalline quality of our samples was confirmed via Raman spectroscopy performed on mechanically exfoliated flakes. The representative Raman spectrum in Fig.1b. exhibits the two characteristic fingerprints of high-quality graphitic systems: an intense G peak at ~1580 cm$^{-1}$ arising from the in-plane sp² C-C bond vibrations, and a sharp, symmetric 2D peak at ~2720 cm$^{-1}$. The single Lorentzian lineshape of the 2D peak, along with the notable absence of the defect-activated D peak, confirms the low defect density and excellent three-dimensional crystalline order of our material, making it an ideal platform for precise optical measurements[35–37]. Optical micrograph of measured flake is presented in Fig.1b.

Before presenting our experimental results, it is crucial to contextualize the motivation for this study by visualizing the significant discrepancies present in established literature. Fig.1c plots the in-plane optical constants — refractive index (n) and extinction coefficient (k) — from several widely cited datasets[29,30,32]. While the overall spectral trends are qualitatively similar, the quantitative values diverge substantially. For instance, in the near-infrared region around 1200 nm, the reported values for the refractive index n differ by as much as 15%. The values for the extinction coefficient k, which dictates material loss, show variations of up to 23% in the same spectral range.

These inconsistencies could present a significant obstacle to the design and predictive modelling of graphite-based photonic and optoelectronic devices. A new, self-consistent dataset is therefore required to address these discrepancies and provide a more accurate basis for future research and device engineering.

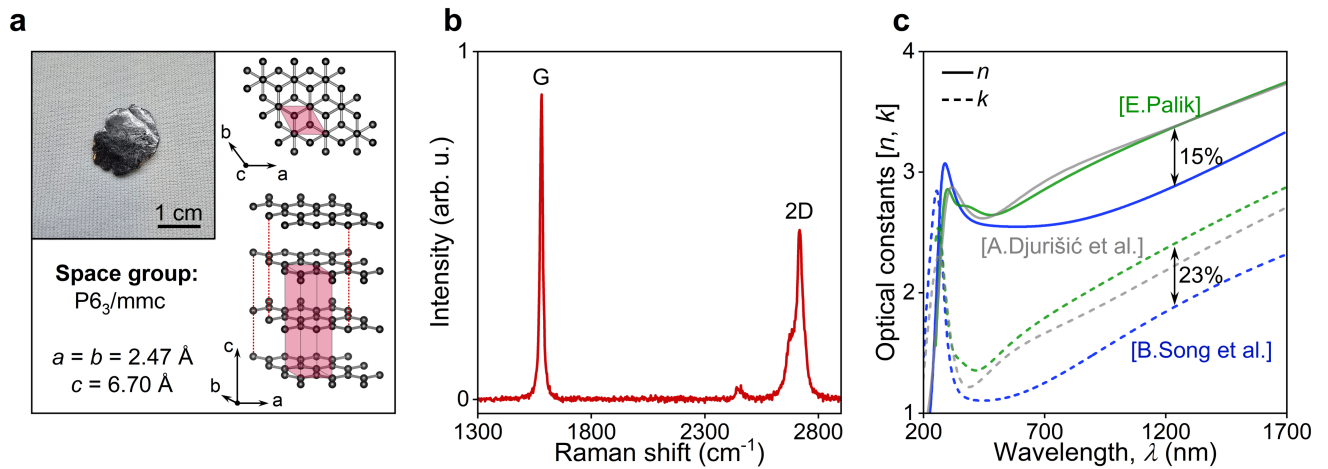

**Fig.1. Structure and optical properties of graphite:** (a) Crystal structure of graphite. The inset represents optical photo of bulk graphite. (b) Raman spectrum of graphite's flake, which is shown at the background of the graph. (c) Comparison of optical constants obtained from different literature data. Solid lines represent refractive index; dashed lines represent extinction coefficient.

**Anisotropic dielectric function**

To address the inconsistencies highlighted previously, we performed variable-angle spectroscopic ellipsometry (SE), a powerful non-destructive technique that is exceptionally sensitive to the optical constants and anisotropy of materials. We measured the ellipsometric angles, Ψ (amplitude ratio) and Δ (phase difference), at four distinct angles of incidence (45°, 50°, 55°, and 60°) over a broad spectral range from 250 nm to 1700 nm (Fig.2a,b). The use of multiple incidence angles is critical, as it provides a set of independent measurements that over-determines the optical system, thereby allowing for the robust and unambiguous extraction of the anisotropic dielectric function.

The raw experimental data (Ψ and Δ) were analyzed using a detailed optical model. Given graphite's hexagonal crystal structure, it is optically described as a uniaxial material with its optic axis oriented along the crystallographic c-axis, perpendicular to the sample's layers. The complex dielectric function is therefore a tensor with two unique components: the in-plane ($\varepsilon_{||} = \varepsilon_{aa} = \varepsilon_{bb}$) and out-of-plane ($\varepsilon_{\perp} = \varepsilon_{cc}$) terms. The resulting real and imaginary parts of the complex dielectric function, $\varepsilon = \varepsilon_1 + i\varepsilon_2$ (or Re[ε] and Im[ε]), are presented in Fig.2c and Fig.2d, respectively. These spectra reveal graphite's profound optical anisotropy.

The in-plane dielectric function ($\varepsilon_{||}$) exhibits a clear metallic character. The real part (Re[ε]$_{||}$) is negative in the ultraviolet (UV) region, crosses zero around 270 nm, and becomes positive at longer wavelengths, a behavior characteristic of a plasmonic material. The imaginary part (Im[ε]$_{||}$), which represents optical absorption, is dominated by two key features. The anisotropy captured by our measurements — the coexistence of metallic and dielectric responses within the same material—is the direct manifestation of the layered electronic structure. Notably, this behavior defines graphite as a natural Type II hyperbolic material in the ultraviolet range (where $Re[\varepsilon]_{||} < 0$ and $Re[\varepsilon]_\perp > 0$), a feature relevant for engineering sub-diffractional light confinement and controlling the local density of optical states.

First, a prominent absorption peak is observed at approximately 280 nm (~4.4 eV), corresponding to the strong π-π* interband transition[38,39]. This single-particle excitation provides the fundamental oscillator strength for the π-plasmon, a collective longitudinal oscillation of the delocalized π electrons. Mediated by long-range Coulomb interactions, this collective mode is shifted to a higher energy, appearing as a distinct peak in the energy-loss function at ~7 eV[11,40].

Second, $Im[\varepsilon]_{||}$ steadily increases toward longer wavelengths in the near-infrared (NIR), which is the signature of Drude-like free-carrier absorption from the semi-metallic band structure of graphite[11,41,42].

In contrast, the out-of-plane component ($\varepsilon_\perp$) displays a classic dielectric (insulating) response. The real part ($Re[\varepsilon]_\perp$) is positive and nearly constant at ~3.6 across the entire spectral range. Correspondingly, the imaginary part ($Im[\varepsilon]_\perp$) is essentially zero, indicating negligible absorption along the c-axis. This extreme anisotropy is a direct consequence of graphite's electronic structure: the high in-plane conductivity is due to the mobile π electrons, while the weak van der Waals forces between layers effectively confine electrons within the planes, leading to insulating behavior in the out-of-plane direction. The self-consistent determination of these two distinct components provides the fundamental basis for accurately calculating the optical constants n and k and resolving the discrepancies found in the literature.

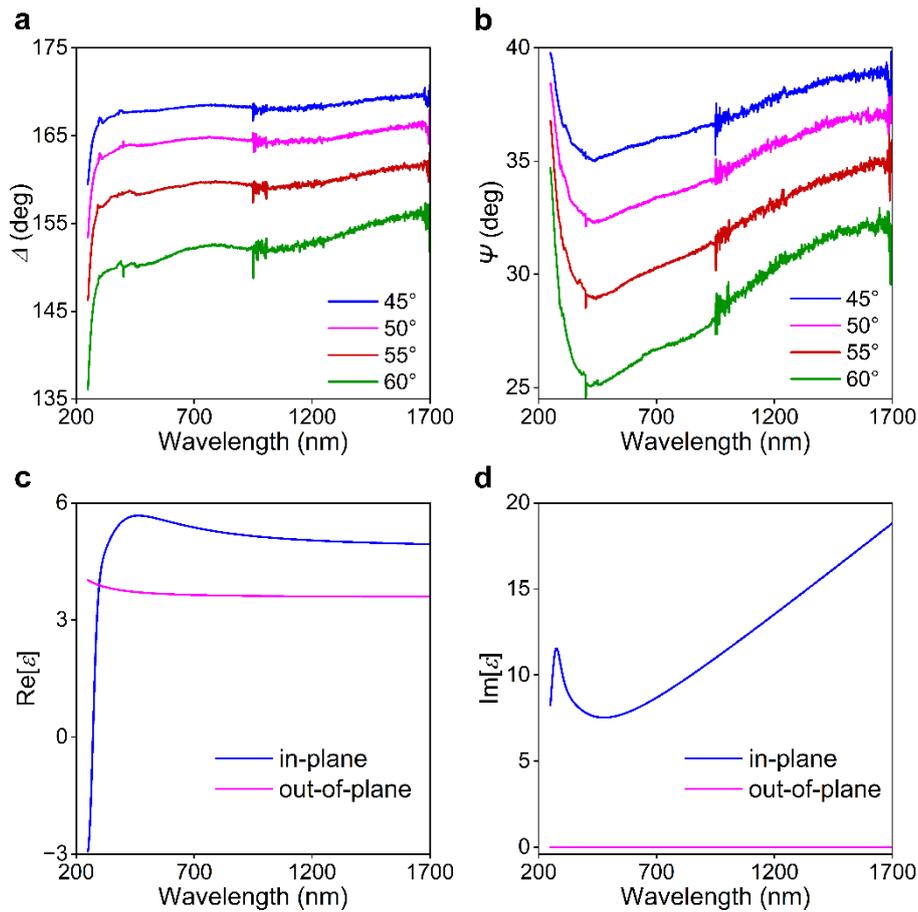

**Fig.2. Ellipsometric parameters:** (a) Δ and (b) Ψ at incident angles 45°, 50°, 55°, 60°. (c) Real and (d) imaginary parts of the dielectric function. In-plane part is depicted by blue line; out-of-plane is presented by magenta line.

**Determination of graphite's optical constants**

The fundamental dielectric function (ε) directly governs the material's phenomenological optical constants — the refractive index (n) and the extinction coefficient (k) — through the relation $\sqrt{\varepsilon} = n + ik$. For interactions with light at or near normal incidence, the response is dominated by the in-plane component of the dielectric tensor. Accordingly, we derived the in-plane optical constants ($n_{||}$ and $k_{||}$) from our experimentally determined dielectric function (Fig.2c,d).

The results are presented in Fig.3, where they are compared to two of the widely referenced datasets in the literature[29,30,32]. Our values for the refractive index (n) are plotted in Fig.2a. The spectrum is characterized by a value of ~2.7 in the green visible region (~550 nm), followed by monotonical increase into the near-infrared (NIR). While the seminal works of E. Palik[32] and A. Djurišić et al.[29] provide a reasonable approximation, our results are systematically lower across the entire spectral range. Conversely, the dataset from B. Song et al.[30] appears to significantly underestimate n, particularly at longer wavelengths. Thus, our findings are situated between the values reported in these prior studies, addressing the discrepancy in the existing literature.

A similar analysis for the extinction coefficient (k) is shown in Fig.3b. The spectrum for k clearly displays the features discussed previously in the context of $Im[\varepsilon]_{||}$: the extinction coefficient is characterized by a sharp peak near 250 nm. This feature is the manifestation of the π-π* interband transition, which corresponds to the peak in the fundamental absorption Im[ε] at ~280 nm. The shift in the peak position between k and Im[ε] is an expected consequence of the Kramers-Kronig relations. Then this peak is followed by subsequent rise in the NIR due to Drude absorption. Our data for k shows good agreement with E. Palik's values in the UV-visible region and A. Djurišić's et al. in NIR. Notably, the data from B. Song et al. deviates significantly, predicting much weaker absorption across most of the spectrum. These quantitative discrepancies underscore the need for a re-evaluation of graphite's optical properties.

To independently verify the accuracy of our ellipsometry-derived dataset, we performed a direct near-normal incidence reflectance measurement on the same sample (Fig.3c). This provides a crucial cross-validation, as reflectance is a direct measure of light intensity and does not rely on the same modelling procedures as ellipsometry. We then used our newly determined n and k values from ellipsometry to calculate the expected reflectance spectrum (R) using the Fresnel equations. As shown in Fig.3c, the calculated spectrum is in excellent agreement with the directly measured experimental spectrum. This strong correspondence between results from two independent optical methods provides confirmation of the accuracy and self-consistency of our determined optical constants, establishing them as a reliable benchmark for future optical design and fundamental studies.

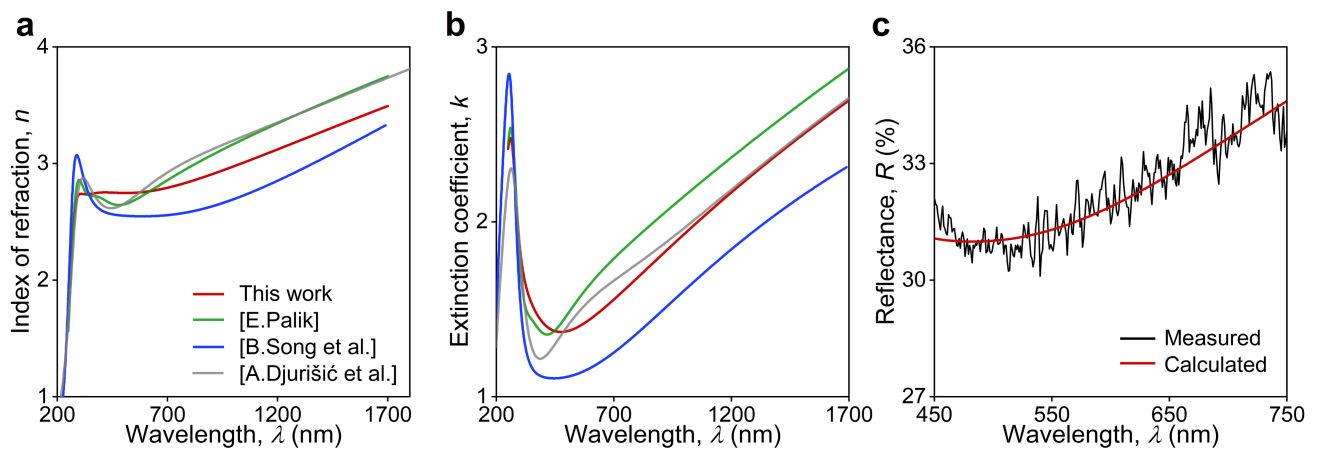

**Fig.3. Comparison of graphite's optical constants:** (a) refractive index, (b) extinction coefficient. (c) Measured and calculated reflectance spectra.

**Near-field measurements of graphite**

To verify the optical constants obtained by ellipsometry and probe graphite's local optical response, we performed scattering-type scanning near-field optical microscopy (s-SNOM) in reflection geometry with pseudo-heterodyne detection (Fig. 4a). This configuration enables simultaneous measurement of amplitude and phase of the scattered field. Illumination and collection were achieved with a parabolic mirror under oblique incidence, and the signal was demodulated at the third harmonic of the tip oscillation to isolate the genuine near-field contribution.

We performed near-field imaging on graphite flakes exfoliated onto $CaF_2$ substrates (Fig. 4b) using infrared excitation (1475–1600 nm). The complex near-field signal (amplitude and phase) from graphite was normalized to that of the $CaF_2$ substrate (Fig. 4c). In this configuration, the near-field interaction is predominantly governed by the out-of-plane dielectric function ($\varepsilon\perp$), particularly for highly anisotropic materials like graphite.

Measurements were conducted on two graphite flakes of 150 nm and 350 nm thickness exfoliated on $CaF_2$ substrates. The sample was illuminated by a tunable laser spanning 1475–1600 nm in 25 nm steps. Each scan included both flake and substrate areas, allowing point-by-point normalization of the complex near-field signal to the $CaF_2$ reference, thereby eliminating instrumental and tip-dependent factors.

Amplitude and phase images were recorded simultaneously (Fig. 4c). The complex near-field signal from graphite was normalized to the response of the $CaF_2$ substrate within each scan. The normalized complex ratio:

$$s_{\rm rel} = \frac{S_{\rm graphite}}{S_{\rm CaF_2}}$$

was analyzed using a dipole-interaction model in which the tip is treated as a metallic sphere interacting electrostatically with the sample[43]. The model parameters correspond to the experimental probe geometry and the optical properties of platinum. The complex permittivity of graphite was numerically retrieved by fitting the modeled and measured $s_{\rm rel}$ (details in Supplementary Note 1).

Crucially, these s-SNOM results are in excellent agreement with the out-of-plane optical constants derived independently from the far-field ellipsometry (solid lines in Fig. 4d). Because the extraction of both in-plane ($\varepsilon_{||}$) and out-of-plane ($\varepsilon_\perp$) components from variable-angle ellipsometry is intrinsically coupled within the uniaxial model, the direct nanoscale validation of $\varepsilon_\perp$ provides strong corroboration for the accuracy and self-consistency of the entire anisotropic dielectric tensor determined in this work. This demonstrates that the macroscopic optical model accurately describes the material's response down to the nanometer length scale.

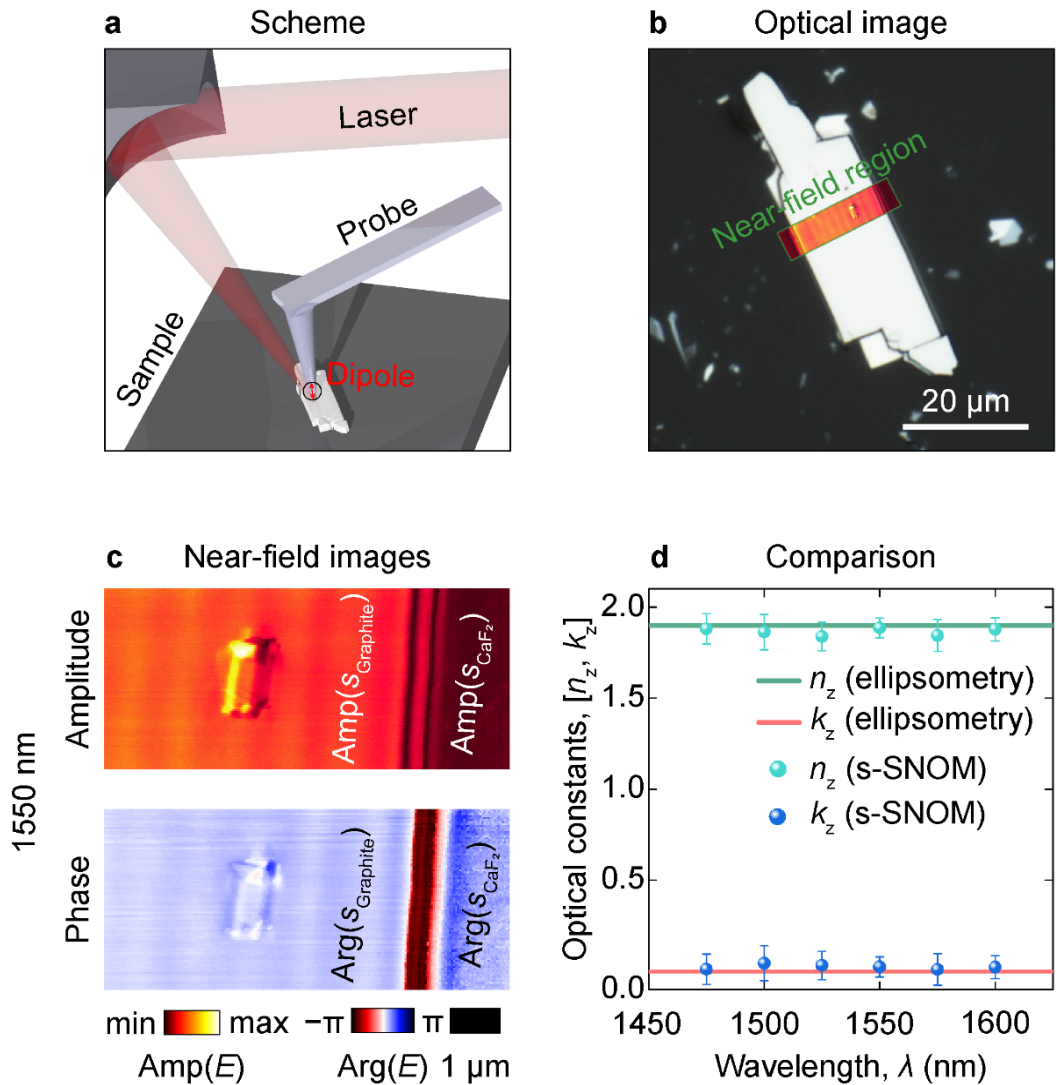

**Fig.4. Near-field extraction of graphite optical constants** (a) Schematic of the reflection-mode s-SNOM experiment. A focused infrared beam is directed and collected by the same parabolic mirror, illuminating a Pt-coated AFM probe that interacts with the sample through a nanoscale dipole field. (b) Optical micrograph of a graphite flake on a CaF$_2$ substrate showing the areas used for atomic-force (AFM) topography and near-field mapping. (c) Representative near-field amplitude (top) and phase (bottom) images acquired at $\lambda$ = 1550 nm. The graphite region exhibits pronounced amplitude enhancement and phase shift relative to the CaF$_2$ substrate, corresponding to differences in local optical response. (d) Spectral dependence of the refractive index $n_z$ and extinction coefficient $k_z$ of graphite obtained from the near-field inversion (symbols) compared with ellipsometry (solid lines).

**From graphene's universal absorbance to the graphite's optical response**

To fully understand describe the graphite's optical properties, we performed a comparative analysis of its optical constants against those of few-layer graphene (Fig. 5). The optical constants, the refractive index and the extinction coefficient, for bulk graphite alongside 1–, 2–, and 3–layered graphene are presented in Fig. 5a,b.

Furthermore, we presented the per-layer absorption, as shown in Fig.5c. In the 2D limit, monolayer graphene exhibits a remarkable frequency-independent optical conductivity, leading to a universal absorbance A = πα ≈ 2.3%[14,15,44–46], a value dictated by the fine-structure constant (α).

As layers are added, the interlayer coupling modifies the electronic band structure, leading to band splitting and a transition towards the semi-metallic band structure of graphite. This electronic evolution is directly reflected in the optical constants (Fig. 5a,b). We observe a systematic increase in both the refractive index and the extinction coefficient with increasing thickness. This trend indicates a renormalization of the electronic structure and enhanced overall oscillator strength as the system evolves towards the bulk limit.

Analyzing the per-layer absorption (Fig. 5c) provides insight into this dimensional crossover. While 1L graphene adheres closely to the universal value πα at longer wavelengths, few-layer graphene and bulk graphite exhibit a convergent behavior in the NIR. However, significant deviations emerge in the UV region, around the π-π* transition. The per-layer absorption in bulk graphite is substantially enhanced compared to monolayer graphene in this spectral region. This enhancement highlights the influence of interlayer interactions and the evolving dimensionality on the collective electronic excitations, underscoring that the optical identity of graphite is distinct from a simple superposition of independent graphene layers.

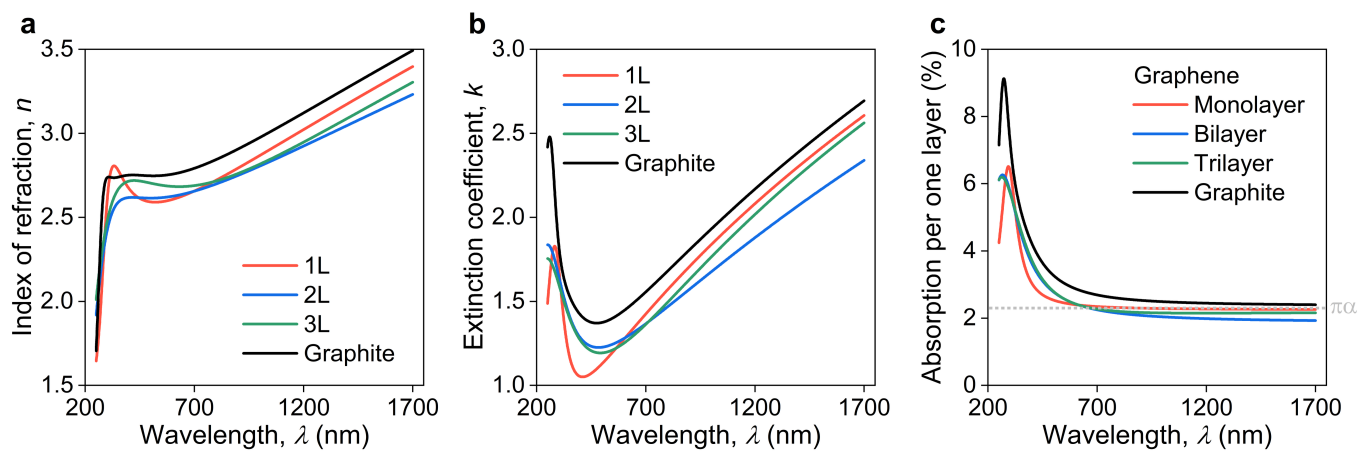

**Fig.5. Optical constants of graphite and 1–, 2–, 3–layered graphene:** (a) refractive index, (b) extinction coefficient. (c) Intrinsic absorbance of graphene.

## Discussion

In conclusion, we have established an accurate, self-consistent dataset for the anisotropic optical constants of bulk graphite. Derived from spectroscopic ellipsometry, these constants are cross-validated by independent micro-reflection spectroscopy and nanoscale near-field measurements, ensuring consistency from macroscopic to nanometer length scales. This multi-technique approach not only provides a new benchmark but also resolves longstanding discrepancies in the literature, which we attribute to material quality and methodological limitations in seminal studies.

The availability of this validated dataset is critical for advancing nanophotonics. It provides the necessary foundation for the high-fidelity design of advanced photonic architectures harnessing the unique anisotropy of graphite, including hyperbolic metasurfaces, polarization-sensitive photodetectors, and complex vdW heterostructures, thereby removing uncertainty in simulations. By establishing graphite as a predictable and reliable nanophotonic platform, our work opens a direct path to accelerating the design and implementation of next-generation optical components.

## Author Contributions

A.T., G.E. and D.G. contributed equally to this work. G.Ts., A.A. and V.S.V. suggested and directed the project. A.T. and N.P. prepared samples. A.T., G.E., D.G., G.T., A.S., N.P., and A.V. performed the measurements and analyzed the data. A.T., G.E., D.G. wrote the original manuscript. All authors reviewed and edited the paper. All authors contributed to the discussions and commented on the paper.

## Competing Interests

The authors declare no competing financial interest.

## Acknowledgements

The authors thank Alexey Tsapenko for the help in measurements.

## Methods

### Sample fabrication

Measured samples have been mechanically exfoliated out of a natural graphite crystal purchased from NGS Trading & Consulting GmbH, Germany (https://www.graphit.de/).

### Atomic force microscopy

Flake thicknesses of graphite were characterized using atomic force microscopy (AFM). Measurements were carried out in tapping mode with a WITec alpha300 RA system equipped with a NanoWorld ARROW-FMR probe (resonant frequency, 75 kHz; spring constant, 2.8 N/m).

### Raman spectroscopy

Raman spectra were acquired at room temperature in a confocal backscattering geometry using a WITec alpha300 RA microscope. A 532 nm laser was used for excitation, with power maintained below 3 mW. The laser beam was focused onto the sample through a 100× objective lens. The collected light was dispersed by a 600 lines/mm grating onto a thermoelectrically cooled CCD detector. Spectra were typically integrated for 5 s with 10 accumulations. All spectral data were processed using WITec Project SIX software.

### Spectroscopic ellipsometry

Ellipsometric measurements were performed via Accurion EP4 (Accurion GmbH) imaging spectral ellipsometer. All measurements were carried in a 250–1700 nm wavelength range at 4 incident angles: 45°, 50°, 55°, 60°.

### Reflection micro spectroscopy

Reflectance spectra were measured in the 450–750 nm range using a setup based on an optical microscope[47,48]. The ocular system was adapted to accommodate a camera (Kiralux 12.3 MP Color CMOS Camera) and a spectrometer (Thorlabs compact spectrometer CCS200), which was connected by an optical fiber (Thorlabs M92L02) with a core diameter of 200 μm. The reflected light was collected from a spot area of <20 μm using a 20X Olympus Plan Achromat Objective with a numerical aperture of 0.4.

**Near-field measurements**

Near-field optical measurements were conducted using a NeaSNOM (Neaspec) in reflection mode with pseudo-heterodyne detection and third-harmonic demodulation. Graphite flakes of 150 nm and 350 nm thickness on $CaF_2$ substrates were examined using Pt/Ir-coated Si tips (radius ≈ 25 nm, oscillation amplitude ≈ 141 nm, frequency ≈ 280 kHz). Illumination was provided by a tunable laser (1475–1600 nm) coupled through a parabolic mirror. Each dataset contained both graphite and substrate regions, allowing direct normalization of the complex near-field signal. Data were processed using the dipole-model inversion (see Supplementary Note 1).

**Numerical calculations of waveguide modes**

Waveguide mode simulations were performed using COMSOL Multiphysics with a 1D slab geometry representing the graphite flake. The experimentally determined ellipsometry-derived refractive index and extinction coefficient were directly used as input parameters. Transverse magnetic (TM) modes from $TM_0$ to $TM_6$ were calculated, with perfectly matched layers (PMLs) applied at the top and bottom boundaries and periodic conditions at the lateral edges. Mode profiles and effective indices were obtained directly from COMSOL simulations.

## References

1. Bernal, J. D. The structure of graphite. *Proc. R. Soc. Lond. Ser. Contain. Pap. Math. Phys. Character* **106**, 749–773 (1997).
2. Popova, A. N. Crystallographic analysis of graphite by X-Ray diffraction. *Coke Chem.* **60**, 361–365 (2017).
3. Novoselov, K. S. *et al.* Electric Field Effect in Atomically Thin Carbon Films. *Science* **306**, 666–669 (2004).
4. Novoselov, K. S. *et al.* Two-dimensional atomic crystals. *Proc. Natl. Acad. Sci.* **102**, 10451–10453 (2005).
5. Novoselov, K. S., Mishchenko, A., Carvalho, A. & Castro Neto, A. H. 2D materials and van der Waals heterostructures. *Science* **353**, aac9439 (2016).
6. Ermolaev, G. A. *et al.* Giant optical anisotropy in transition metal dichalcogenides for next-generation photonics. *Nat. Commun.* **12**, 854 (2021).
7. Grudinin, D. V. *et al.* Hexagonal boron nitride nanophotonics: a record-breaking material for the ultraviolet and visible spectral ranges. *Mater. Horiz.* **10**, 2427–2435 (2023).
8. Niu, S. *et al.* Giant optical anisotropy in a quasi-one-dimensional crystal. *Nat. Photonics* **12**, 392–396 (2018).
9. Slavich, A. S. *et al.* Exploring van der Waals materials with high anisotropy: geometrical and optical approaches. *Light Sci. Appl.* **13**, 68 (2024).
10. Cong, C., Shang, J., Wang, Y. & Yu, T. Optical Properties of 2D Semiconductor WS2. *Adv. Opt. Mater.* **6**, 1700767 (2018).
11. Taft, E. A. & Philipp, H. R. Optical Properties of Graphite. *Phys. Rev.* **138**, A197–A202 (1965).
12. Lin, M. F., Huang, C. S. & Chuu, D. S. Plasmons in graphite and stage-1 graphite intercalation compounds. *Phys. Rev. B* **55**, 13961–13971 (1997).
13. Vitali, L. *et al*. Phonon and plasmon excitation in inelastic electron tunneling spectroscopy of graphite. *Phys. Rev. B* **69**, 121414 (2004).

14. Nair, R. R. *et al.* Fine Structure Constant Defines Visual Transparency of Graphene. *Science* **320**, 1308–1308 (2008).
15. Toksumakov, A. N. *et al.* Anomalous optical response of graphene on hexagonal boron nitride substrates. *Commun. Phys.* **6**, 13 (2023).
16. Craciun, M. F., Russo, S., Yamamoto, M. & Tarucha, S. Tuneable electronic properties in graphene. *Nano Today* **6**, 42–60 (2011).
17. Fernández-Rossier, J., Palacios, J. J. & Brey, L. Electronic structure of gated graphene and graphene ribbons. *Phys. Rev. B* **75**, 205441 (2007).
18. Weiss, N. O. *et al.* Graphene: An Emerging Electronic Material. *Adv. Mater.* **24**, 5782–5825 (2012).
19. Oostinga, J. B. *et al*. Gate-induced insulating state in bilayer graphene devices. *Nat. Mater.* **7**, 151–157 (2008).
20. Grigorenko, A. N., Polini, M. & Novoselov, K. S. Graphene plasmonics. *Nat. Photonics* **6**, 749–758 (2012).
21. Low, T. & Avouris, P. Graphene Plasmonics for Terahertz to Mid-Infrared Applications. *ACS Nano* **8**, 1086–1101 (2014).
22. Mak, K. F. *et al.* Atomically Thin MoS2: A New Direct-Gap Semiconductor. *Phys. Rev. Lett.* **105**, 136805 (2010).
23. Xiao, D. *et al*. Coupled Spin and Valley Physics in Monolayers of MoS2 and Other Group-VI Dichalcogenides. *Phys. Rev. Lett.* **108**, 196802 (2012).
24. Kadantsev, E. S. & Hawrylak, P. Electronic structure of a single MoS2 monolayer. *Solid State Commun.* **152**, 909–913 (2012).
25. Chaves, A. *et al.* Bandgap engineering of two-dimensional semiconductor materials. *Npj 2D Mater. Appl.* **4**, 29 (2020).
26. Peng, Z. *et al*. Strain engineering of 2D semiconductors and graphene: from strain fields to band-structure tuning and photonic applications. *Light Sci. Appl.* **9**, 190 (2020).
27. Kormányos, A. *et al.* k·p theory for two-dimensional transition metal dichalcogenide semiconductors. *2D Mater.* **2**, 022001 (2015).
28. Zotev, P. G. *et al.* Van der Waals Materials for Applications in Nanophotonics. *Laser Photonics Rev.* **17**, 2200957 (2023).
29. Djurišić, A. B. & Li, E. H. Optical properties of graphite. *J. Appl. Phys.* **85**, 7404–7410 (1999).
30. Song, B. *et al.* Broadband optical properties of graphene and HOPG investigated by spectroscopic Mueller matrix ellipsometry. *Appl. Surf. Sci.* **439**, 1079–1087 (2018).
31. Greenaway, D. L., Harbeke, G., Bassani, F. & Tosatti, E. Anisotropy of the Optical Constants and the Band Structure of Graphite. *Phys. Rev.* **178**, 1340–1348 (1969).
32. Palik, E. D. *Handbook of Optical Constants of Solids*. (Academic Press, 1998).
33. Papoular, R. *et al*. The vis/UV spectrum of coals and the interstellar extinction curve. *Astron. Astrophys.* **270**, L5–L8 (1993).
34. Smausz, T. *et al.* Determination of UV–visible–NIR absorption coefficient of graphite bulk using direct and indirect methods. *Appl. Phys. A* **123**, 633 (2017).
35. Ferrari, A., Robertson, J., Reich, S. & Thomsen, C. Raman spectroscopy of graphite. *Philos. Trans. R. Soc. Lond. Ser. Math. Phys. Eng. Sci.* **362**, 2271–2288 (2004).
36. Tene, T. *et al.* Drying-Time Study in Graphene Oxide. *Nanomaterials* **11**, 1035 (2021).
37. Boroujerdi, R., Abdelkader, A. & Paul, R. Highly Sensitive and Selective Detection of the Antidepressant Amitriptyline Using a Functionalised Graphene-Based Sensor. *ChemNanoMat* **8**, e202200209 (2022).
38. Marinopoulos, A. G., Reining, L., Rubio, A. & Olevano, V. Ab initio study of the optical absorption and wave-vector-dependent dielectric response of graphite. *Phys. Rev. B* **69**, 245419 (2004).
39. Kinyanjui, M. K. *et al.* Direct probe of linearly dispersing 2D interband plasmons in a free-standing graphene monolayer. *Europhys. Lett.* **97**, 57005 (2012).
40. Eberlein, T. *et al.* Plasmon spectroscopy of free-standing graphene films. *Phys. Rev. B* **77**, 233406 (2008).

41. Ashcroft, N. W., & Mermin, N. D. *Solid State. Physics*. vol. 1 (1976).
42. Fox, M. *Optical Properties of Solids*. vol. 3 (Oxford university press, 2010).
43. Cvitkovic, A., Ocelic, N. & Hillenbrand, R. Analytical model for quantitative prediction of material contrasts in scattering-type near-field optical microscopy. *Opt. Express* **15**, 8550–8565 (2007).
44. Kravets, V. G. *et al.* Spectroscopic ellipsometry of graphene and an exciton-shifted van Hove peak in absorption. *Phys. Rev. B* **81**, 155413 (2010).
45. Matković, A. *et al.* Spectroscopic imaging ellipsometry and Fano resonance modeling of graphene. *J. Appl. Phys.* **112**, 123523 (2012).
46. Weber, J. W., Calado, V. E. & van de Sanden, M. C. M. Optical constants of graphene measured by spectroscopic ellipsometry. *Appl. Phys. Lett.* **97**, 091904 (2010).
47. Frisenda, R. *et al.* Micro-reflectance and transmittance spectroscopy: a versatile and powerful tool to characterize 2D materials. *J. Phys. Appl. Phys.* **50**, 074002 (2017).
48. Slavich, A. *et al.* Optical Properties of Biaxial van der Waals Crystals for Photonic Applications. *Bull. Russ. Acad. Sci. Phys.* **88**, S433–S438 (2024).